\documentclass[]{aastex631}

\begin{document}

\def\RSUN{R$_{\sun}$ }
\def\MSUN{M$_{\sun}$ }
\def\kms{$\rm km~s^{-1}$}
\def\kmss{$\rm km~s^{-1}$ }
\def\cmcube{$\rm cm^{-3}$}
\def\cmcubes{$\rm cm^{-3}$ }
\def\arcs{\char'175\ ~}
\def\arcsc{\char'175 }
\def\arcm{\char'023\ ~}
\title{A Nitrogen-rich SNR in M31: SNR Interaction with the CSM at Late Times }


\correspondingauthor{Nelson Caldwell}
\email{caldwell@cfa.harvard.edu}

\author[0000-0003-2352-3202]{Nelson Caldwell}
\affiliation{Center for Astrophysics, Harvard \& Smithsonian, 60 Garden Street, Cambridge, MA 02138, USA}
\author[0000-0002-7868-1622]{John Raymond}
\affiliation{Center for Astrophysics, Harvard \& Smithsonian, 60 Garden Street, Cambridge, MA 02138, USA}
\date{May 2023}



\begin{abstract}
We present the discovery of a supernova remnant (SNR) in M31 which is unlike any other remnant known in that galaxy. An optical ground-based spectrum of WB92-26 taken at the MMT and sampling most of this marginally resolved object reveals strong lines of [O II], [Ne III], H I, [O III], [O I], [N II] and [S II], though the H I lines are very weak and the [N II] lines are very strong. 
Multiple velocity components are visible in those lines, with broad wings extending to  $-2000$ and $+1500$ or $2000$ \kmss (the heliocentric velocity of M31 is $-300$ \kms).  The lines show strong peaks or shoulders near $-750$ \kms, $-50$ \kms, and $+800$ \kmss in the M31 frame.

The density implied by the [S II] ratio combined with the X-ray luminosity, FUV flux and optical size lead us to conclude that the optical emission lines are generated by shock waves, not photoionization. Consideration of the velocity structure indicates that the emission is from a shock in the circumstellar medium (CSM). This CSM must be depleted in H and enriched in He and N through CNO processing, and it must have had a high velocity before the explosion of the parent star, to explain the broad wings in the emission lines. We estimate the CSM shell to have a mass of 2 M$_\sun$, implying a Core Collapse SN.  It is likely that Eta Car will produce a remnant resembling WB92-26 a few thousand years after it explodes.

\end{abstract}


\section{Introduction} \label{sec:intro}

Supernova remnants (SNRs) that show optical emission from SN ejecta and circumstellar medium (CSM) are rare, but they provide unique information about the SN progenitor and about the explosion. There are a handful of O-rich SNRs in the Galaxy and the Magellanic Clouds such as Cas A, G292.0+1.8, N132D and 1E102-72.3 \citep{hwang12, temim22, blair00}, 
but only three O-rich SNRs are known in other galaxies \citep[NGC~4449 and M83,][]{blair84, lee14, winkler17}, and
it is somewhat surprising that so few SNRs like Cas A have been found in surveys of other galaxies. N-rich SNRs are even more rare.  Some small knots in Cas A show strong [N II] emission \citep{alarie14}, and the N:O ratio in the X-ray and IR emitting gas of RCW 103 is about 3.5 times the solar value \citep{millard21, narita23}.  Possible nitrogen overabundance in Kepler's SNR has been attributed to mass loss from an evolved companion \citep{chiotellis12}, though the overabundance of N relative to O has not been firmly established \citep{dopita19}. 

Surveys of SNRs in M31 have been conducted in the optical by \citet{blair82, braun93, lee14}, the radio by \citet{dickel84}, the X-ray by \citet{stiele11, sasaki12, sasaki18} and the UV by \citet{leahy23}.  Here we discuss a supernova remnant in M31 that was not included in  these catalogs, yet shows an underabundance of H and overabundance of He and N indicative of CNO processing.  The emission extends from $-2000$ \kmss to $+1500$ \kmss ($-1700$ to $+1800$ in the frame of M31), indicating that it is a young SNR.  The ratios of O, Ne and S to H are also high, apparently due to conversion of H to He in the CNO process.  This SNR is therefore among the rare SNRs in other galaxies that show strong nuclear processing.


The subject of this paper was observed as part of spectroscopic study of star clusters and emission line sources in the Andromeda galaxy, as observed with Hectospec/MMT \citep{fab} which was described in \cite{C09} (C09) and \cite{sanders}, though our object was not discussed in either of those papers.  The observed objects in Sanders et al. were chosen to have strong H$\alpha$ or [OIII] but to be nearly unresolved on narrow band images \cite[the imaging material was from the Local Group Galaxy Survey, LGGS,][] {M06}, being the signatures of either PNe or a compact  H\,{\sc ii} region. SNRs were not expected to appear because of the apparent size limitation imposed. Other M31 emission-line projects such as the early 
\cite{WB} (WB) and \cite{BW} papers and the more recent \cite{lee14} (LL) and \cite{AZ} projects also involved narrow band images centered on H$\alpha$ and/or [SII] and cataloged discrete emission-line objects, the latter two also using the LGGS data. The WB and LL projects searched in particular for targets that had strong [SII] compared to H$\alpha $ (greater than 0.4), as a means of finding SNRs. Our object was cataloged as an M31 emission-line object in WB, but was not included in their subsequent SNR paper, ostensibly because its [SII]/H$\alpha $ ratio was too low for a SNR.  Nevertheless, we use the WB92-26 designation for the object in preference to our own name, as WB were the first to publish the source (we note that it is designated "WB92a 26" in SIMBAD, (\cite{simbad}). An additional Hectospec/MMT spectroscopic study of fainter PNe appears in \cite{bhatt} and another of symbiotic stars also using LGGS as source material \citep{miko} with the latter chosing discrete targets that had H$\alpha $ emission accompanied by red broad-band colors indicative of the presence of a late-type star. The three Hectospec emission-line surveys were all successful in finding the intended classes of emission line objects. However, while compiling the first survey in 2009 (C09), the first author found that a single observed target had a spectrum with complex velocity structure at H$\alpha$ with strong [SII], and made a brief note that the target was likely a SNR. No further analysis was done for quite some time. More recently, while producing a catalog of all of the extant M31 Hectospec studies for public archive purposes, the spectrum was inspected visually again, where it became clear that it was unique of all the 8600 M31 targets observed by Hectospec/MMT, even including the often complex WR and LBV spectra obtained by \cite{Massey2} also with Hectospec, because of the aforementioned velocity structure of WB92-26. This review made another fact clear, the complex structure seen at $\sim 6560$\AA \ is mostly due to [NII], not hydrogen, for  H$\beta $ is weak and the same velocity complexity is also apparent in the other optical region forbidden lines. The weak H$\beta $ line cannot be explained by high reddening of the target because [OII]$\lambda 3727$ is strong, thus H$\alpha$ must be only a small contributor to the emission at 6560\AA , meaning of course that [NII] is strong. We came to the conclusion that the reason the target does not appear in any of the catalogs produced by previous SNR searches using narrow-band images such as WB is that the ratio of [SII] to emission at 6560\AA \ is lower than 0.4, artificially so because forming such a ratio using narrow-band images for this object is effectively a ratio of [SII]/[NII] and not [SII]/H$\alpha$.

\begin{figure}
\begin{centering}
\includegraphics[width=7.0in]{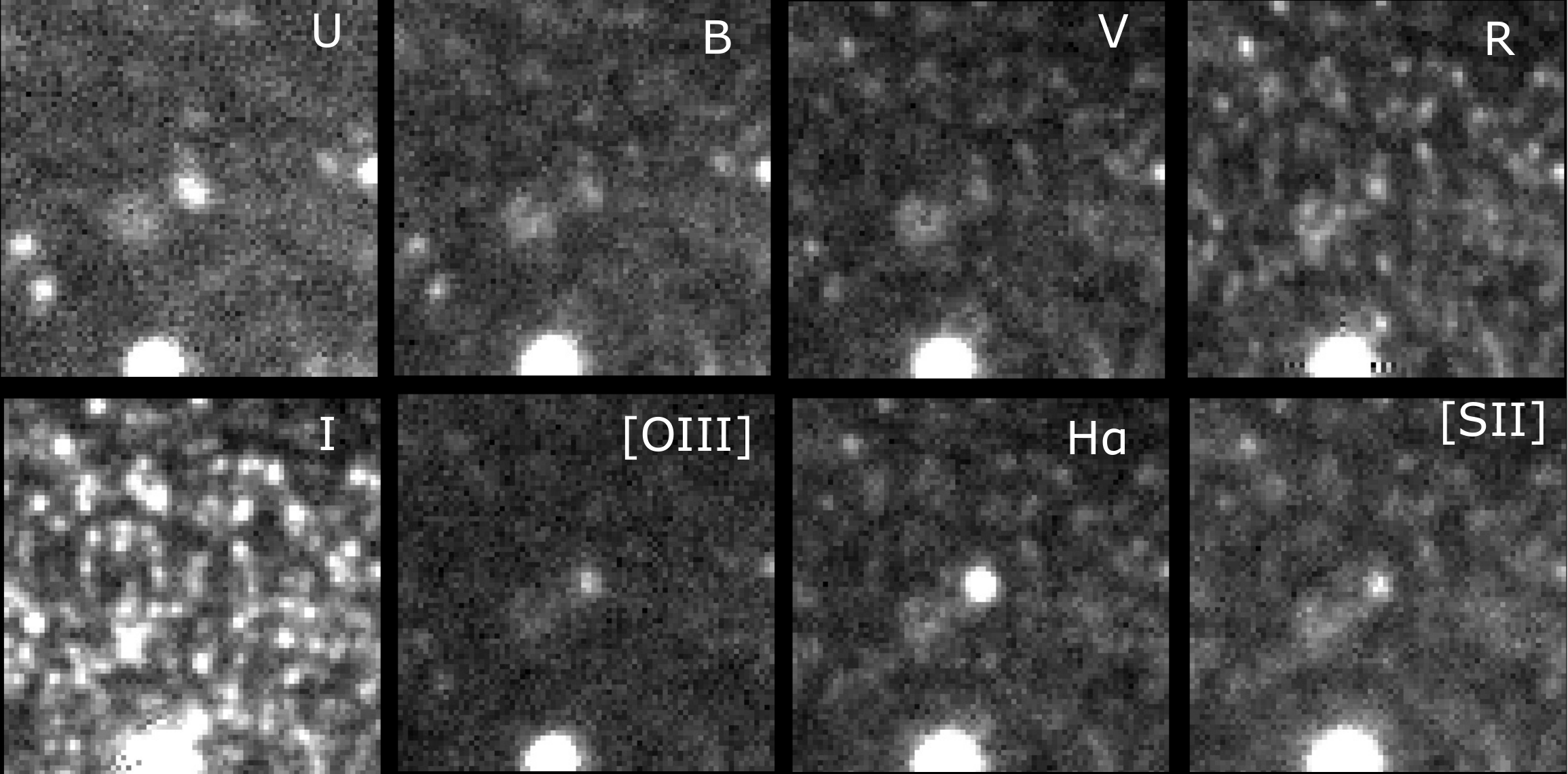}
\caption{Ground-based images of W92-26 taken as part of the LGGS survey \citep{M06}. Shown are UBVRI broad-band filter images, each covering 20\arcs (75 pc), and narrow-band images in [O III], H$\alpha $ region, and [S II]. All images in this figure and Fig. \ref{where} have north at the top and east to the left. The SNR is barely seen in the redder broad-band images, but is visible in the U-band probably because of the presence of [O II]$\lambda 3727$ emission, where the object is slightly resolved.
\label{images}
}
\end{centering}
\end{figure}

\begin{figure}
\begin{centering}
\includegraphics[width=3.0in]{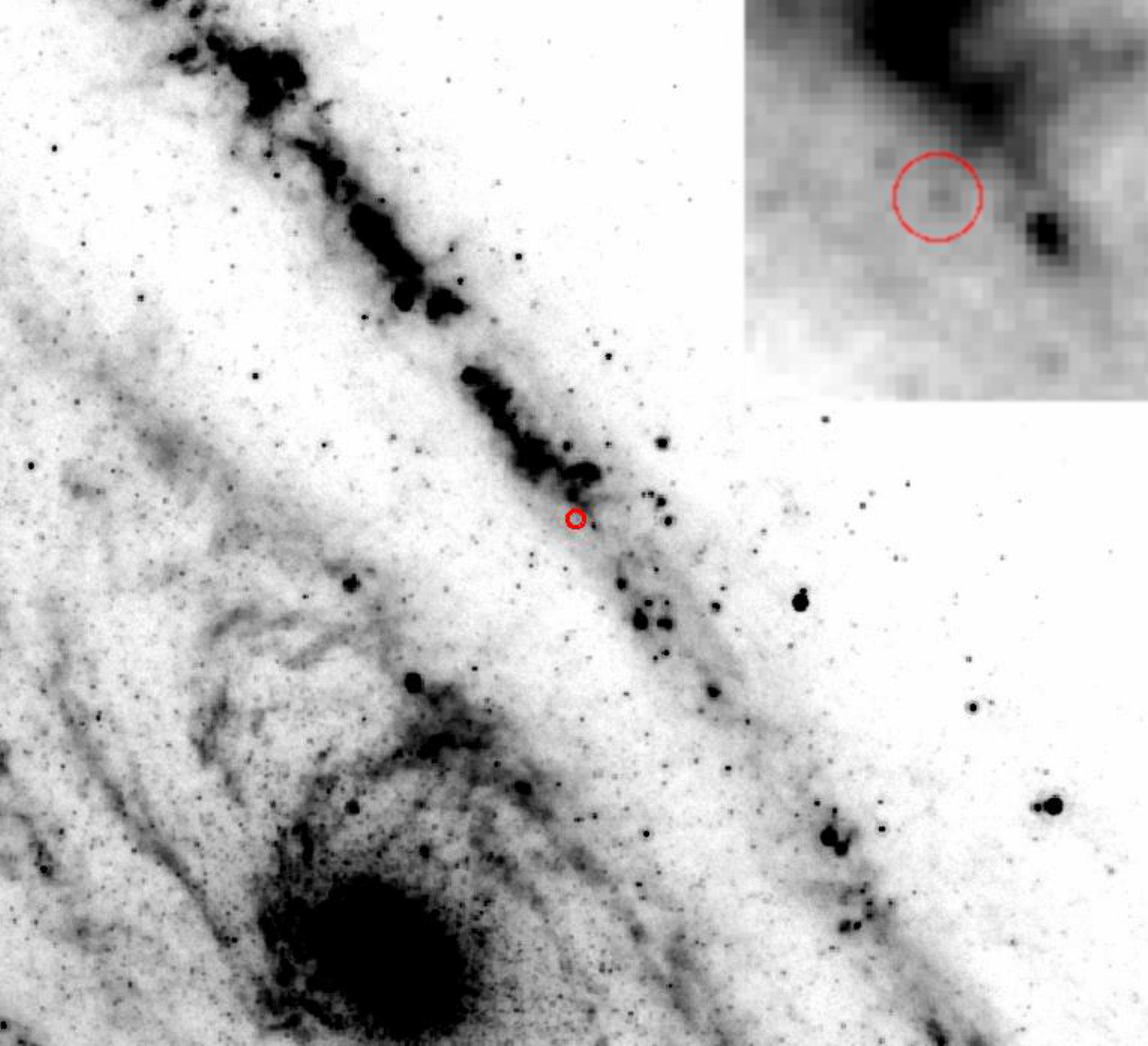}
\caption{Spitzer MIPS $24\mu $\ image \citep{Gordon} showing the position of WB92-26 in the M31 galaxy (open red circle, centered on the spectroscopic coordinates), which is located in the 10 kpc star formation ring (seen here as the arc from top left to bottom right). For reference, the M31 bulge is at the lower left. The image shown is 25\arcm on a side (5.7 kpc). The inset image is 1.5\arcm across (340 pc).
\label{where}
}
\end{centering}
\end{figure}

WB92-26 is nearly  unresolved in LGGS images as expected; Figure \ref{images} shows the available LGGS images. We measure an H$\alpha $\ FWHM of 1.2\arcsc, noting that star images are 0.9\arcs. The full diameter at zero-intensity level is 3.2", or about 12 pc for an M31 distance of 773 kpc \citep{Conn}. The U-band image shows the source to be somewhat larger and asymmetric, with a FWHM of 1.5\arcs compared to nearby stars with 1.0\arcsc. There is no HST image of the area, nor is it part of the projected observing plan. Figure \ref{where} shows the position of the object in M31, amidst a portion of the ``10 kpc" star formation ring \citep{habing} as clearly seen in this Spitzer $24\mu $ MIPS image \citep{Gordon}. In the study of X-ray point sources in the M31 bulge using the Chandra HRC, \cite{kaaret} detected WB92-26 (it is item 22 in his list) and gave a flux of $6.9 \pm 3.4 \times 10^{-6} $ photons cm$^{-2}$ s$^{-1}$, and a luminosity of $1.7 \times 10^{36}$ erg s$^{-1}$.
It was also detected by XMM (and given the label 2XMM $J004220.2+412640$) by \cite{stiele11}, with a flux of $5.12 \times 10^{-15}$ erg cm$^{-2}$ s$^{-1}$. 

The mean velocity derived from the [OIII] line discussed below is $343 \pm 5$ \kms, a value that is reasonably close to the velocity expected from the \cite{kent} M31 rotation model at that position ($-300$ \kms), given that the object is an SNR.  






The paper is organized as follows: Section 2 describes the observations and data analysis, Section 3 discusses the interpretation in terms of shock waves and the CSM, and Section 4 summarizes our conclusions.

\section{Observations}

\begin{figure}
\begin{centering}
\includegraphics[width=5.0in]{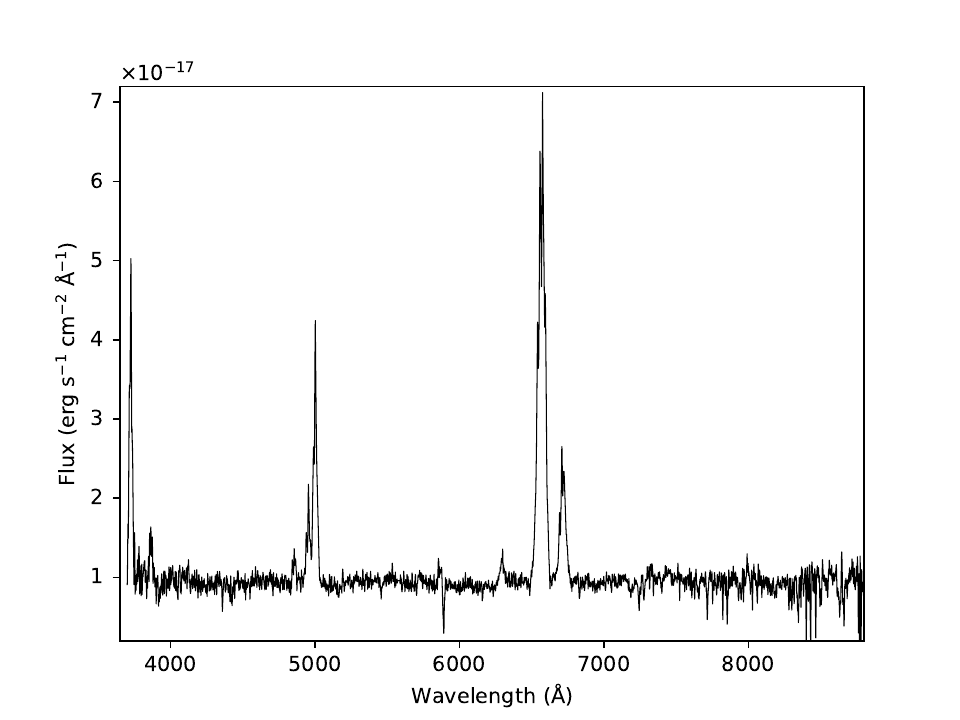}
\caption{Full spectrum of WB92-26 taken with the MMT/Hectospec. Total exposure time was 14400s. The continuum light is due to stars not associated with the SNR.
\label{spectrum}
}
\end{centering}
\end{figure}

WB92-26 was observed as part of three separate Hectospec fields, done on 2006 Oct 23, 2006 Nov 17 and 2007 Oct 21. The fiber spectra, which cover a circle of 1.5\arcs on the sky, were reduced in a standard manner, as described in C09. The resultant spectra have $\lambda/\Delta\lambda = 1000$. For the first two dates, measurements 10\arcs from the targets were also made, with the goal of producing spectra with a local background subtracted instead of just the nighttime sky, the latter typically created from spectra of blank fields nearer to the edge of the 1 degree field, and thus away from the disk of M31. However, the S/N of the spectra of this object are better if indeed just the more distant nighttime sky spectrum is used. Thus we elected to use the sum of all three exposures, none of which were corrected for the local background. To be certain, we have also inspected the two local background-subtracted spectra and can confirm there is little continuum at the position of the SNR. Thus in
particular the NaD absorption seen in Fig. \ref{spectrum}, is due to the starlight surrounding the SNR. The spectra from the three observations were added, giving a total exposure time of 14400 seconds. This spectrum is available at the MMT/Hectospec M31 archive (\href{https://oirsa.cfa.harvard.edu/signature\_program/}{https://oirsa.cfa.harvard.edu/signature\_program/}) by searching either on its SIMBAD name or its coordinates. Flux standards observed during the observing runs, though not at the particular nights, were used to provide a relative flux correction. The MMT/Hectospec instrument has been shown to have stable throughput corrections over time \citep{fab}. The flux zero-point could be set by using the total H$\alpha $ luminosity tabulated in \cite{AZ}, though this does require us to assume that the applied extinction of A$_{\rm V} = 1.412$ is correct, which as we discuss below is probably too high.

\begin{figure}
\begin{centering}
\includegraphics[width=5.0in]{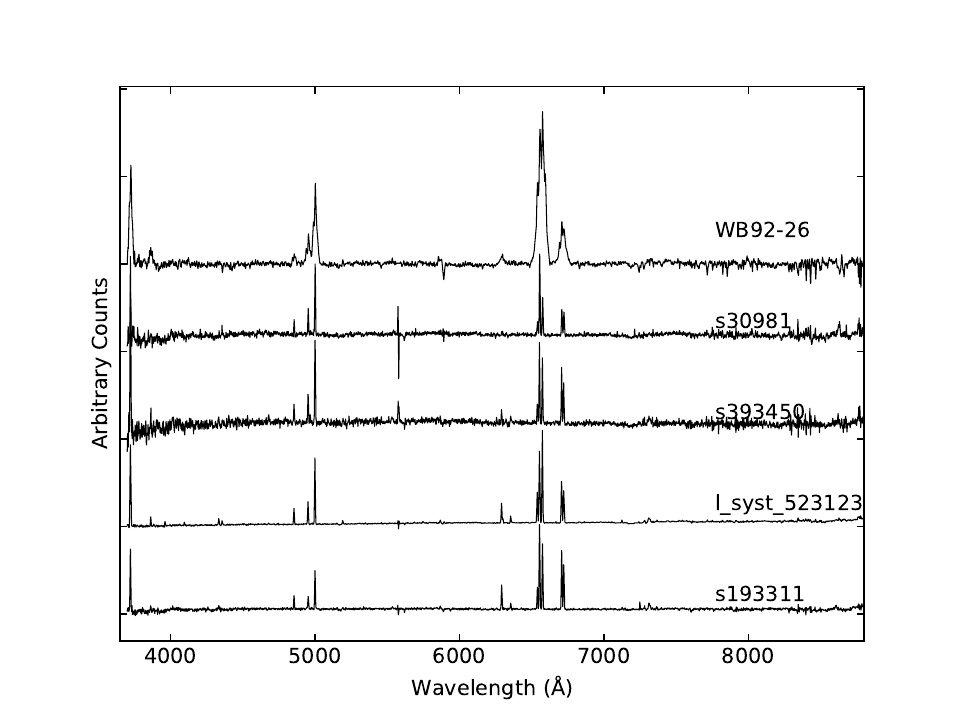}
\caption{Full spectrum of four other SNRs in M31 as comparisons for WB92-26, shown at the top. These were also taken with MMT/Hectospec, and were selected to be shown here because of they all have high [SII]/H$\alpha$ ratios and were also detected in Xrays. These optical spectra are typical of SNRs, yet very different from the spectrum of WB92-26.
\label{more_spectra}
}
\end{centering}
\end{figure}

Figure \ref{spectrum} shows the three co-added spectra. Lines of [O II], [Ne III], H I, [O III], [O I], [N II], [S II] are seen, with broad  wings extending to $-2000$ and $+1500$ or $2000$ \kmss.  They show strong peaks or shoulders near $-750$ \kmss and $0$ \kms.  The [N II] lines show a peak near $+800$ \kms, while the other lines show shoulders on the emission profiles near that velocity, but not distinct peaks. This spectrum is very unusual among SNR spectra, both in its line widths and in its relative line fluxes.  Table \ref{line_ratios} and Figure \ref{more_spectra} show spectra of four other SNRs also observed with Hectospec.  Emission from these ordinary SNRs comes from $\sim$100-200 \kmss shocks in the ISM.  Their line widths are at most marginally resolved, and the line ratios are typical of shocks in the ISM of our own galaxy.

\begin{deluxetable*}{lllrrrrll} 
\tablecolumns{9}
\tablewidth{0pt} 
\tablecaption{Line Ratios\label{line_ratios}} 

\tablehead{\colhead{Name}& \colhead{RA} & \colhead{DEC} &\colhead{[O III]/H$\beta$} & \colhead{[N II]/H$\alpha$} & \colhead{[S II]/H$\alpha$} & \colhead{FWHM\tablenotemark{a}} & \colhead{cflux\tablenotemark{b}} & \colhead{Note\tablenotemark{c}}\\
\hline
\colhead{} &\colhead{} &\colhead{} &\colhead{} &\colhead{} &\colhead{} &\colhead{\AA} &\colhead{ergs s$^{-1}$ cm$^{-2}$ }  &\colhead{}\\
}
\startdata 
\hline
WB92-26      & 10.5851292 & 41.4446411 & 13.03 & 11.75 & 2.69 & 18.52 & 1.5e-15 &\\
l\_syst\_523123& 10.3984167 & 41.1150094 & 4.09  & 1.26  & 1.03 & 5.37 & 1.4e-15 &LL \\  
s393450      & 10.5448333 & 40.8634339 & 5.05 &  0.810  & 1.17 & 5.17 & 7.8e-16 &\\
s193311      & 10.5610417 & 40.8681258 & 2.89 &  0.75  & 1.20  & 4.82 & 6.0e-16 &
\\
s30981       & 10.7305417 & 40.9962578 & 5.55 &  0.51 &  0.66 & 5.31 & 5.9e-16 &LL\\ 
\enddata 
\tablenotetext{a}{Measured using the [OIII] $\lambda 5007$ line. The instrumental resolution is 5\AA, so all
of the widths except for that of WB92-26 are unresolved.}
\tablenotetext{b} {Combined EPIC flux from \cite{stiele11}}
\tablenotetext{c} {Also listed as SNR in \cite{lee14}}
\end{deluxetable*}

\begin{figure}
\begin{centering}
\includegraphics[width=5.0in]{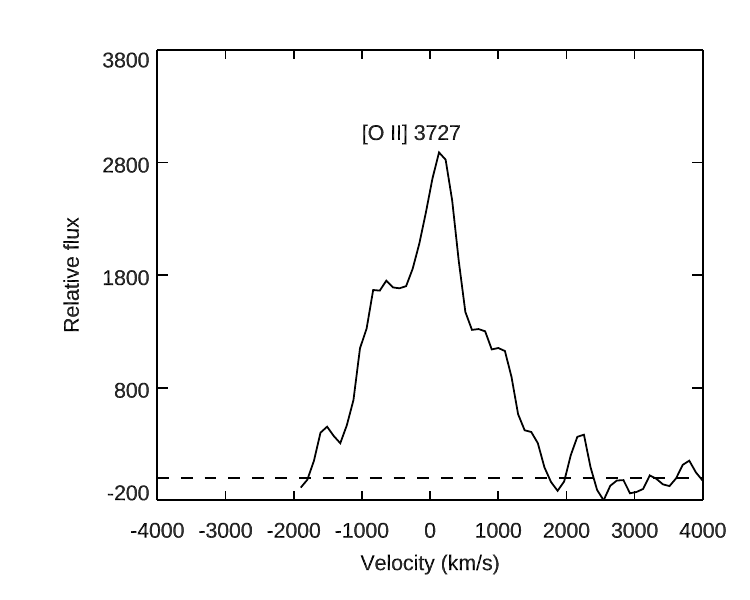}
\caption{Profile of the [O II] doublet emission, showing peaks near -700 \kmss and +200 \kms. The velocity scale pertains to the $\lambda$3728.91 line relative to the -300 \kmss velocity of this part of M31. Because [O II] is at the edge of the spectral range, the continuum level is poorly defined.
\label{o2_unfolded}
}
\end{centering}
\end{figure}

\begin{figure}
\begin{centering}
\includegraphics[width=5.0in]{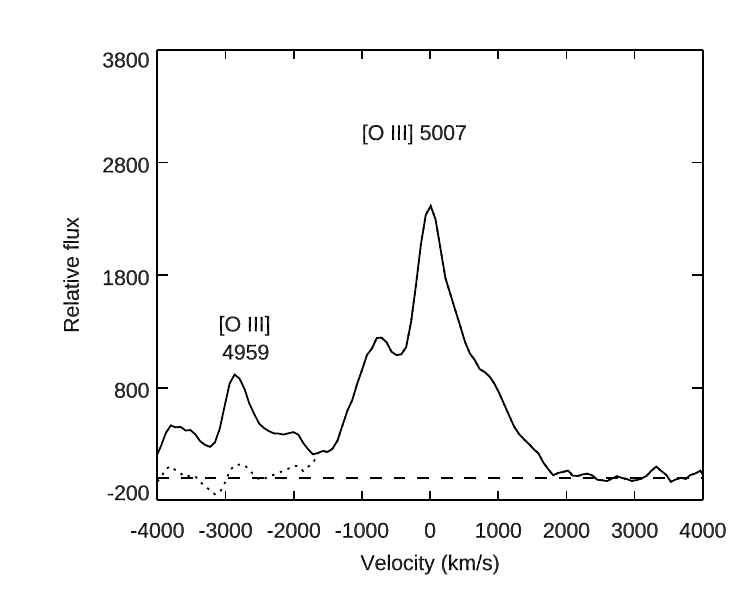}
\caption{Profile of the [O III] doublet emission on the velocity scale based on the $\lambda$5006.84 line at the -300 \kmss velocity of this part of M31.  It shows peaks near -700 \kmss and 0 \kmss and a shoulder near +850 \kms.  The blend of $\lambda$4959 and $\lambda$5007 emission was unfolded as described in the text.  The solid line is the observed profile and the dotted line is the $\lambda$5007 contribution.
\label{o3_unfolded}
}
\end{centering}
\end{figure}

\begin{figure}
\begin{centering}
\includegraphics[width=5.0in]{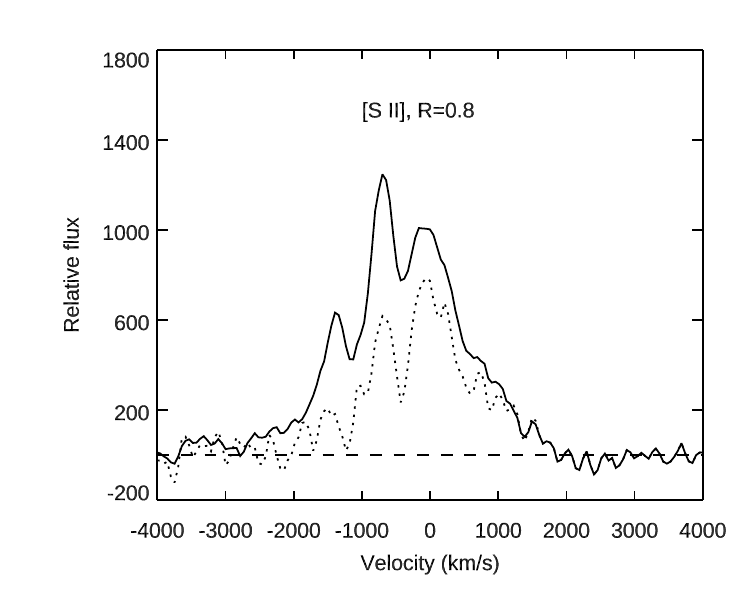}
\caption{Profile of the [S II] doublet emission on the velocity scale of the $\lambda$6730.78 line at the -300 \kmss velocity of this part of M31.  The blend of $\lambda$6717 and $\lambda$6731 emission was unfolded as described in the text, with a line separation of 640 \kmss and an assumed ratio of $\lambda$6717 to $\lambda$6731 intensities of 0.8.  The solid line is the observed profile and the dotted line is the $\lambda$6731 contribution.  The $\lambda$6717 profile shows peaks at about -750 \kmss and 0 \kms, with a shoulder near +750 \kms.
\label{s2_unfolded}
}
\end{centering}
\end{figure}

\begin{figure}
\begin{centering}
\includegraphics[width=5.0in]{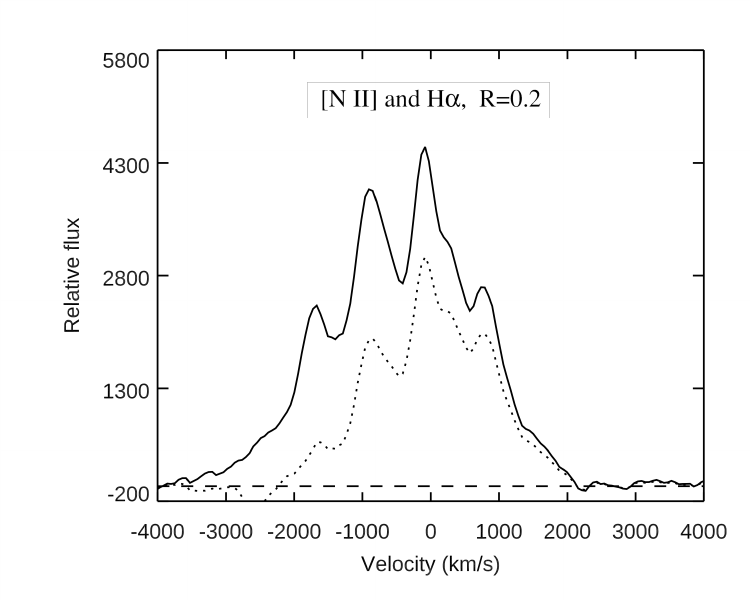}
\caption{Profile of the [N II] doublet emission on a scale based on the $\lambda$6583.4 line at the $-300$ \kmss velocity of this part of M31.  It shows peaks near $-850$ \kms, $-100$ \kmss and $+800$ \kms, along with a secondary peak at +300 \kms.  The blend of $\lambda$6548 and $\lambda$6584 emission was unfolded as described in the text, with a line separation of 1620 \kmss and an assumed ratio of H$\alpha$ to $\lambda$6584 intensities of 0.2.  The solid line is the observed profile and the dotted line is the $\lambda$6584 contribution.
\label{n2_unfolded}
}
\end{centering}
\end{figure}

Figures~\ref{o2_unfolded}-\ref{n2_unfolded} show the profiles of the [O II], [O III], [N II]+H$\alpha$ and [S II] lines.  All of these are complicated by their multiplet structures, so we have attempted to unfold them.  For [O III], the ratio of intensities of the $\lambda$$\lambda$4959, 5007 lines is set by the Einstein A values of the $^1D_2$ level at 1:3, and their wavelength separation is equivalent to a velocity difference of 2880 \kms.  Therefore, we start at the red edge of the profile, where the $\lambda$5007 emission must dominate, and step toward the blue, subtracting the $\lambda$4959 contribution by subtracting 1/3 of the flux at each pixel from the flux of the pixel 2880 \kmss to the blue.  The solid line in Figure~\ref{o3_unfolded} is the observed profile, and the dotted line is the $\lambda$5007 contribution.  This procedure makes no assumptions about symmetry or clumpiness of the emitting gas.  It only assumes the 3:1 intensity ratio for $\lambda$$\lambda$5007, 4959 and the known wavelength separation of the lines.  We have checked the procedure by comparing the resulting 5007 and 4959 profiles.  We also tried by starting at the blue and of the spectrum and subtracting 3 times the 4959 emission from the position in the blended profile 2880 \kmss to the red.  That test produces a similar profile, but it is noisier because it multiplies any errors by a factor of 3, rather than dividing them by 3 when the subtraction starts with 5007 at the red end of the profile.  The separation of the [O II] doublet at 3728.9 and 3726.2 is only 220 \kms, and the doublet is not resolved, so we have not attempted to unfold the contributions.  The [O II] and [O III] profiles both show peaks at roughly -800 \kms\/ and 0 \kms , along with shoulders near +850 \kms.  However, because the [O II] profile is noisier than the [O III] profile and less well-resolved, it would be dangerous to interpret the more detailed similarities or differences of the profiles.

For [S II], the separation between the $\lambda$6731 and $\lambda$6717 lines is 640 \kms, but the intensity ratio depends on the density.  We assumed various intensity ratios between the high- and low-density limits and followed a procedure like that used for [O III]. Figure~\ref{s2_unfolded} shows the unfolded 6731 profile assuming that the 6717/6731 ratio is 0.8.  There is no reason to expect that the ratio is constant over the whole profile, so there is some ambiguity about the resulting profile.  However, a ratio of 0.8 gave the greatest similarity between the 6717 and 6731 profiles, and it does not produce systematically negative fluxes or fluxes above the observed values for either line, while 6717/6731 ratios smaller than 0.7 or larger than 0.9 give problematic results.  Therefore, we take the ratio to between 0.7 and 0.9 and the density to be 1000 to 3000 \cmcube\/ based on ratios predicted by CHIANTI version 8 \citep{delzanna15}.  The peak in the observed profile near -1400 \kmss is really 6717 at about -750 \kms, the peak near -750 \kmss has comparable contributions from the two lines, and the peak near 0 \kmss is dominated by 6732.  

The [N II] $\lambda$6548 to $\lambda$6584 ratio, like the [O III] ratio, is fixed at 1/3, but there is also a contribution from H$\alpha$.  Figure~\ref{n2_unfolded} follows the same procedure as [O III], but with a separation of 1620 \kmss between the [N II] lines, and it assumes that H$\alpha$/$\lambda6584$ is 0.20 everywhere in the profile.  One might expect a smaller ratio of 0.10 from the H$\beta$ flux and the usual Balmer decrement of H$\alpha$/H$\beta$=3, but that leads to negative $\lambda$6584 fluxes near -2500 \kms.  The explanation for this discrepancy could lie in our assumption that the H$\alpha$/6584 ratio is constant across the profile, in the calibration of the spectrum, in reddening, or in a departure of the Balmer decrement from the value usually seen in SNRs.  It is difficult to reconcile even moderate reddening with the ratio of [O II]$\lambda$3727 to the upper limit on [O II]$\lambda$7325, and while slightly steeper Balmer decrements are observed in nonradiative SNR shocks, that emission is faint \citep{ghavamian02}.  Therefore, the first option, that the unfolding is not entirely accurate, is the most probable, and we take H$\alpha$/6584 to be 0.1-0.2. That range is mainly constrained by the H$\beta$ flux, the Case B Balmer decrement, and the range of reddenings discussed below.

With the help of the unfolding described above, we measure the line fluxes.  Table \ref{line_fluxes} presents the observed fluxes relative to H$\beta$=100 along with the fluxes corrected for reddenings of E(B-V)=0.1, which is near the Galactic reddening toward M31 \citep[0.06,][]{schlafly} and  for E(B-V)=0.4, which corresponds to the extinction A$_V$=1.412 assumed by \citet{AZ}.  The red wing of the He I $\lambda$5876 line is blended with Na I absorption, so we have only a lower limit to its flux.  Several weaker He I lines should accompany the $\lambda$5876 line, and the one at $\lambda$4473 appears to be present, but the others are blended with much stronger lines. The [O II] multiplet at $\lambda$7325 appears to be marginally present, but we can be confident only of an upper limit.  The reddening to WB92-26 is not really known. The Balmer decrement based on the H$\alpha$ intensity from the unfolded blend of [N II] and H$\alpha$ suggests E(B-V) $\simeq$0.4, but that is uncertain.  On the other hand, a reddening significantly higher than 0.4 would lead to very high [O II] $\lambda$$\lambda$7325/3727 ratios larger than observed.  The higher reddening would also be incompatible with the UV emission apparently detected by \citet{leahy20} (See section 3).  Therefore, we favor E(B-V)$\sim$0.1.  We do not quote uncertainties because systematic uncertainties dominate.  The transmissions of individual fibers vary by about 10\%.  The relative fluxes of the lines should be accurate to the calibration uncertainty of about 10\% and the uncertainties in the unfolding of blends.  

We can now finish the discussion above regarding the early mis-classification of WB92-26. With the effective removal of [N II] from the H$\alpha $ area we can form a proper [SII]/H$\alpha $ ratio, finding the ratio to be 0.5 for the E(B-V)=0.1 case, and 0.55 for the 0.4 case, both quite different from the upper limit ratio 0.26 found using narrow-band imaging \citep{WB}. The new ratio trivially indicates the spectrum is that typical of an SNR.

\section{Interpretation}

\subsection{Density}
The [S II] doublet ratio of 0.7-0.9 indicates a density
of 1000-3000 \cmcube.  That density is high enough to modestly suppress the [O II] $\lambda$3727 emission and increase [O II] $\lambda$7325.  Densities much above 1000 \cmcube\/ would drive the 7325/3727 ratio above the upper limit in Table \ref{line_fluxes}.  We therefore take 1500 \cmcube\/ to be the density in the optical emission region.    This density would strongly suppress the [N I] $\lambda5200$ doublet, which has a critical density near 200 \cmcube.  That is in agreement with the [N I]/[O I] ratio. 

\subsection{Excitation mechanism}

The emission could be produced by shock waves, as in most SNRs, but it might also be produced by photoionization, as in the Crab \citep{sibley16}.  In this case, we combine the X-ray luminosity of $1.7\times10^{32}~\rm erg~s^{-1}$ from \citet{kaaret02}, the density of 1500 \cmcube\/ and the SNR radius of 5 pc to obtain the ionization parameter $\xi$=L/(nr$^2$) = $10^{-5}$.  That is far too low to produce the observed ionization state \citep{kallman82}, and we conclude that the optical emission lines are generated by shock waves.   

\subsection{Composition}

The most obvious feature of the spectrum shown in Figure~\ref{spectrum} and Table \ref{line_fluxes} is the unusual strength of all the lines relative to the Balmer lines.  He I $\lambda$5876 is more that 0.74 times as strong as H$\beta$, while both models and observations of SNR shocks in normal abundance gas show the ratio to be $<$0.25 \citep{hartigan87, raymond20}.  Similarly, [O III]/H$\beta$ is about 3 times the values typical of SNR shocks, and [N II]/H$\alpha$ is about 9 times the typical values.   This pattern of abundances indicates that the gas has been through some CNO processing.

Given that the emission arises from shocks in the SN ejecta or CSM, the Hectospec fiber would include light from a broad range of shock speeds, preshock densities and elemental abundances.  Therefore, no single shock model should be expected to accurately represent the spectrum.  However, we can get some idea of the average abundances from a shock model.   Table \ref{line_fluxes} includes a model based on the current version of the code described by \citet{raymond79}.  It assumes a shock speed of 150 \kms, a preshock density of nucleons (n$_H$+4n$_{He}$) of 5 \cmcube\/ and a preshock magnetic field of 0.5$\mu$G.  The abundances compared to solar abundances assume that about 3/4 of the H has been burned to He and the some of the carbon has been transformed to nitrogen, oxygen and neon.  The assumed abundance set is H:He:C:N:O:Ne:Mg:Si:S = 12.0: 12.2: 8.8: 9.3: 9.7: 9.0: 8.4: 8.5: 8.1.  We assume that the electron and ion temperatures are equal, as is found for shocks of modest Mach number \citep{ghavamian13, medina14, raymond23}.  This model differs from the models of shocks in ejecta composed entirely of metals \citep{itoh81, sutherland95, blair00, raymond18} because the plasma is dominated by H and He, so the radiative cooling rate is not large enough to drive the electron temperature below the ion temperature. 

The table shows that this simple model does a reasonable job of matching the observed fluxes within the uncertainty in reddening.  However, a single model does not match both the highest ionization lines ([O III] and [Ne III]) and the lowest ([N I] and [O I]) simultaneously.  In addition to the fact that a range of shock speeds is present in the SNR, there is probably a contribution from a photoionization  \citep{sutherland95, docenko10} or a cosmic ray \citep{hester94} precursor. Overall, we take the agreement to mean that the abundance set listed above is a reasonable average for the emitting region, though we have no direct constraints at all on C or Si.

The model assumes that the elements are evenly mixed within the emitting region, but observations of Cas A, for example, show strong abundance variations among the small optical knots.  For instance, \citet{koo23} find clusters of knots that are bright in He, S, or Fe. On larger scales the X-ray emitting gas shows regions bright in elements such as Si or Fe \citep{hwang12}.  The velocity profiles of the strong lines in WB92-26 are very similar except that the +800 \kmss peak in [N II] only appears as a modest shoulder in [O III] and [S II].  The similarity is consistent with even mixing of the material, but variations on smaller scales are certainly possible.  The profiles of the weaker lines, such as H$\beta$ and [O I] are too noisy to allow a clear  statement about their similarity to the profiles of the other lines. 

WB92-26 is not well resolved in groundbased images, and the M31 point source catalog based on AstroSat/UVIT images of \citet{leahy20} lists a source 1.8\arcsec\/ from WB92-26 with an FUV flux of $8.24 \times 10^{-15}~\rm erg~cm^{-2}~s^{-1}$ in the CaF2 band (1233-1733\AA). \citet{AZ} use a reddening A$_V$=1.41 to estimate the (H$\alpha$+[N II]) luminosity of the nebula based on the Spitzer maps of \citet{Gordon}.  The shock model shown in Table \ref{line_fluxes} predicts UV fluxes 85\% the observed value for E(B-V)=0.0 or 50\% the observed value for E(B-V)=0.1, but an order of magnitude too low for the value E(B-V)=0.4 adopted by \citet{AZ}.  Comparing other sources in the UVIT catalog near our object with ground-based optical coordinates, there may be a small bulk coordinate offset.  We consider it probable that the object in the \citet{leahy20} catalog is WB92-26, that the model in Table \ref{line_fluxes} applies to this object to within a factor of 2, and that the reddening is smaller than that assumed by \citet{AZ} because the object is located on the near side of the plane of M31. If the UV flux does indeed originate from WB92-26, it argues strongly for shock excitation rather than photoionization, since shocks produce the high temperatures needed for efficient excitation of UV lines.


\subsection{Ejecta or CSM?}

Given that the emission originates in shock waves, it is necessary to understand the velocity structure.  While a shock seems like the simple explanation given the 1500 \kmss velocities observed, shocks faster that around 700 \kmss will heat the gas to over 10$^7$ K, and they will not have time to cool back down to produce the observed optical lines.  Depending on the age of the SNR and the density of the shocked material, shocks below about 500 \kmss are needed to produce the optical spectrum.  If the material is SN ejecta, that can be accomplished if the layer of ejecta with speeds of 1500-2000 \kmss is now encountering a reverse shock whose speed is a few hundred \kmss in the dense ejecta.  As in Cas A, slow shocks in dense knots could produce the optical emission, while faster shocks in the interclump ejecta could produce X-rays.  That could occur if the SN ejecta are moving in a low density bubble created by a progenitor wind.  However, equating the ram pressures of the $\sim$150 \kmss shocks in a density of 5 \cmcube\/ that produce the optical emission with the $\sim$1500 \kmss shock in the bubble implies a bubble density of 0.05 \cmcube.  Combining that with the SNR volume, the expected X-ray luminosity is two orders of magnitude below the observed value.  Therefore, we conclude that the optical emission probably comes from shocks in the CSM.

In the CSM picture, the requirement is again that the gas be moving at $\sim$1500 \kms, but that the shock speed be of order 150 \kms. That will occur if the CSM is moving rapidly, and its inner surface is hit by lower density ejecta.  The CSM observed in many SNe and in Cas A is only moving at tens to hundreds of \kmss \citep{chevalier78, milisavljevic15, weil20, jacobson-galan20}.  However, a high velocity CSM is seen in the case of Eta Carinae. The N-rich material in the lobes of Eta Car has velocities of order 650 \kms, and other N-rich material is seen at higher speeds \citep{davidson82, gull20}.  Therefore an SN explosion in a system like Eta Car could plausibly produce the abundance pattern and line profiles seen in WB92-26.  Nitrogen enhancement is seen in the circumstellar shells of some SNRs.  The quasi-stationary flocculi (QSFs) and the photoionized CSM in Cas A are enriched in He and N by an order of magnitude \citep{chevalier78, weil20}.  RCW 103 shows strong enhancement of N in the X-ray shell \citep{narita23}, as well as in the ISO infrared spectrum, where a 900 \kmss wide line of [N II] 122$\mu$ is the brightest line in the far IR \citep{millard21}.  \citet{narita23} interpret the approximately 4 times overabundance of N relative to O in RCW 103 as an indication that the X-ray emission comes from shocked CSM material, and the 900 \kmss line width may indicate that the CSM had a high velocity before the explosion.  If the SN ejecta overtake the CSM at a more modest speed, such as 500 \kms, and if the ejecta are strongly enriched in metals, the X-ray luminosity could be comparable to the observed value.

Overall, we conclude that the optical emission of WB92-26 could originate from shocks in a high velocity shell of CNO-processed circumstellar shell analogous to that of Eta Car.  Eta Car shows two symmetric expanding lobes expanding at $\pm$650 \kms, which might account for the peaks in the WB92-26 line profiles near +800 and -750 \kms. The lobes of Eta Car will take about 5,000 years to reach the 5 pc radius of WB92-26.  A limitation of the analogy is that Eta Car has a WR star companion, and its interaction with the primary creates the double-lobe structure.  However, a similar massive WR star that survived the explosion would probably be visible in our spectrum.  This might put a limit on the mass of the companion.

\begin{deluxetable*}{lrrrrr} 
\tablecolumns{5}
\tablewidth{0pt} 
\tablecaption{Line fluxes relative to H$\beta$=100 \label{line_fluxes}}
\tablehead{\colhead{Ion}& \colhead{$\lambda$} & \colhead{F} &\colhead{I$_{0.1}$} & \colhead{I$_{0.4}$} & \colhead{Model}\\}
\startdata 
\hline
O II    & 3727 &  989 & 1096 & 1512 &  750 \\
Ne III  & 3864 &  203 & 222 & 297  &  92 \\
H$\beta$  & 4861 &  100 & 100 & 100 & 100 \\
O III   & 4959 &  329 & 324 & 317 & 178 \\
O III   & 5007 &  975 & 959 & 923 & 534 \\
N I     & 5199 &   34 & 33  & 30 & 41 \\
He I\tablenotemark{a}  & 5876 &$>$97&$>$90 &$>$74 & 120 \\
O I     & 6300 &  133 & 121 &  93 & 79  \\
N II    & 6548 &  899 & 811 & 600 & 280  \\
H$\alpha$ & 6563 &  450 & 406 & 300 & 300 \\
N II    & 6584 & 2690 & 2424 & 1801 &  840 \\
S II    & 6717 &  361 & 323 & 235  & 219 \\
S II    & 6731 &  451 & 404 & 293  & 257 \\
O II    & 7325 &  $<$55& $<$48 & $<$32 & 51 \\
\\
\hline
H$\beta\tablenotemark{b}$  &  & 0.21 & 0.26 & 0.81  &  \\
\hline
\enddata 
\tablenotetext{a}{The red wing of the He I line is swallowed by unrelated Na I absorption.}
\tablenotetext{b}{ $10^{-16} ~\rm erg~cm^{-2}~s^{-1}$ in the 1.5\arcsec\/ diameter fiber.  F, I$_{0.1}$ and I$_{0.4}$ are the observed flux and fluxes corrected for E(B-V)=0.1 and 0.4, see text.} 
\end{deluxetable*}

\subsection{Overall Structure}

The shock model gives a surface flux of I(H$\alpha$+N II) = 9.1$\times 10^{-4}~\rm erg~cm^{-2}~s^{-1}$ of shock front with a preshock density n$_0$ of 5 nucleons (H+He) \cmcube .  For a radius of 5 pc, the area is given by a covering factor f$_1$ times $2\times 10^{39} \rm ~cm^2$ .  The observed H$\alpha$+[N II] luminosity of 2.7$\times 10^{35}~\rm erg~s^{-1}$ (assuming E(B-V)=0.1) then implies that $n_0 f_1$ = 0.7 \cmcube.  If the shell is 5  pc in radius and 1 pc thick, that amounts to about 2 \MSUN\/ of CSM.  The [S II] ratio implies an electron density of order 1500 \cmcube\/ in the recombination zone downstream.  In the model shown in the table, this is density reached with with a modest preshock density and a very low magnetic field, as might be expected in SN ejecta or CSM.  Higher preshock densities and magnetic fields could also give [S II] ratios near the observed value, but then a covering factor farther below 1 would be required.  The high estimated mass suggests a core-collapse SN.

\section{Summary}

WB92-26 is a young SNR whose optical emission is dominated by N-rich gas that has undergone enough CNO processing to convert roughly 3/4 of its hydrogen to helium.  We do not have definitive evidence that the gas is CSM rather than ejecta, but it is difficult to understand the X-ray luminosity in the ejecta picture.  Therefore, the picture that most naturally explains the observed optical spectrum and X-ray luminosity is that a massive star exploded inside a circumstellar medium much like the CSM of Eta Car.  

Eta Car is a binary system containing a $\sim$100 \MSUN primary and a WR star \citep{smith18}.  The CSM of Eta Car is quite complex \citep{morse01, gull20}.  The interactions within the binary system and a series of eruptions 
have produced a double-lobed structure called the Homunculus that expands at about 650 \kmss embedded in a lower density, higher velocity patchy wind.  An equatorial disk, a ``jet" and knots called Weigelt clumps are also seen.  The gas is highly nitrogen enriched \citep{davidson82, smith04} as a result of CNO processing.  The total mass of the CSM is estimated to be 10-45 \MSUN \citep{smith03, morris17}, even larger than our very rough estimate for WB92-26.  In principle, the CSM might extend beyond the current size of the SNR, which would give a larger mass.   

There is no evidence for a companion star left behind in the explosion of WB92-26. The broad-band images show no likely candidate and as stated above, the spectrum with local sky-subtraction shows no continuum remaining. Also, there are certainly no strong emission lines typical of WR stars \citep[C~IV or N~V,][]{neugent}.

About 5,000 years from now, the Homunculus of Eta Car will have reached a radius of about 3 pc and the more diffuse gas will have reached 5 pc.  If Eta Car explodes then, the shock will reach that material perhaps a thousand years later, and if the CSM is substantially denser than the ejecta, shocks of order 150 \kmss will be driven into the CSM.  A double-lobed dense structure could give rise to the +800 and -750 \kmss peaks observed in WB92-26, while shocks in the more diffuse gas could produce the smooth line wings extending to 1500 or 2000 \kms.  Faster shocks in the ejecta would produce X-rays.  The CSM around Eta Car is quite dusty, so one would expect some IR emission, perhaps explaining the knot of dust emission seen in the Spitzer map of WB92-26  (Fig. \ref{where}).  Thus, while we cannot exclude other scenarios, this one naturally explains the main features of the observation.



A question is begged by the discovery of this SNR: are there any other similarly compact remnants with a complicated velocity structure in M31 that remain to be observed?  The basic discovery data of WB92-26 shows that it has a small size (FWHM less than 1.2\arcs or 5 pc), shows strong emission in the H$\alpha$ \ region (though most of that is actually due to [NII]), and was detected by the XMM survey of \cite{stiele11}. An artificially low [SII]/H$\alpha$ \ ratio as measured from narrow-band images could be checked later via spectroscopy as we have done here.  
 
To investigate the question, we correlated the coordinates of the XMM data of \cite{stiele11} with the H$\alpha $ \ discrete source catalog of \cite{AZ} and the SNR catalog of \cite{lee14}.
Most of the correlations have sizes much too large to be interesting, but
there are about 40 objects in both \cite{stiele11} and  \cite{AZ}  with diameters smaller than 30 pc, 13 of which were observed spectroscopically, including the 5 SNR shown here. Of those not observed, only one is of possible interest with the rest clearly being   H\,{\sc ii} regions or parts of large SNRs. There may be other compact remnants without X-ray detections which would not be counted in this analysis, but those are likely of no interest.

The catalogs prepared for the MMT/Hectospec observations reported in C09, \cite{sanders},  \cite{miko} and \cite{bhatt} resulted in 4000 compact emission-line targets observed with Hectospec/MMT. It was expected that all of those would be PNe, normal  H\,{\sc ii} regions or symbotic stars, but about 50 have proved to have SNR-like [SII]/H$\alpha $ \ emission-line ratios. Even WB92-26 is flagged as strange in a BPT diagram using [SII]/H$\alpha $ versus [NII]$\lambda 6583$/H$\alpha $ as the criteria, proving the necessity of spectroscopy over imaging in such work. There appear to be several hundred other compact emission-line sources that await spectroscopic observations in hopes of finding another object like WB92-26, but only one of these has detectable X-ray emission. In general the current large spectroscopic data set argues that WB92-26 is a unique SNR in M31.

Further observations of WB92-26 would be very fruitful.  HST images could give an accurate size and morphology.  UV spectra from COS would provide the C and Si abundances and a constraint on the shock speeds from the N V/N III] and N V/N IV] ratios.  Models of the evolution of massive stars including the ejection of a massive CSM and the interaction between the supernova and the CSM could constrain the parameters of the CSM, the ejecta and the time between the CSM ejection and the explosion. 





\begin{acknowledgments}
We thank Paul Harding and Heather Morrison for help with producing the original observing catalogs.
\end{acknowledgments}

%

\vspace{5mm}
\facilities{MMT(Hectospec)}


\software{CHIANTI \citep{delzanna15},
IRAF \citep{iraf}
          }


:

\bibliography{WB92-26}{}

\begin{thebibliography}{}
\expandafter\ifx\csname natexlab\endcsname\relax\def\natexlab#1{#1}\fi
\providecommand{\url}[1]{\href{#1}{#1}}
\providecommand{\dodoi}[1]{doi:~\href{http://doi.org/#1}{\nolinkurl{#1}}}
\providecommand{\doeprint}[1]{\href{http://ascl.net/#1}{\nolinkurl{http://ascl.net/#1}}}
\providecommand{\doarXiv}[1]{\href{https://arxiv.org/abs/#1}{\nolinkurl{https://arxiv.org/abs/#1}}}

\bibitem[{{Alarie} {et~al.}(2014){Alarie}, {Bilodeau}, \& {Drissen}}]{alarie14}
{Alarie}, A., {Bilodeau}, A., \& {Drissen}, L. 2014, \mnras, 441, 2996,
  \dodoi{10.1093/mnras/stu774}

\bibitem[{{Azimlu} {et~al.}(2011){Azimlu}, {Marciniak}, \& {Barmby}}]{AZ}
{Azimlu}, M., {Marciniak}, R., \& {Barmby}, P. 2011, \aj, 142, 139,
  \dodoi{10.1088/0004-6256/142/4/139}

\bibitem[{{Bhattacharya} {et~al.}(2019){Bhattacharya}, {Arnaboldi}, {Hartke},
  {Gerhard}, {Comte}, {McConnachie}, \& {Caldwell}}]{bhatt}
{Bhattacharya}, S., {Arnaboldi}, M., {Hartke}, J., {et~al.} 2019, \aap, 624,
  A132, \dodoi{10.1051/0004-6361/201834579}

\bibitem[{{Blair} {et~al.}(1982){Blair}, {Kirshner}, \& {Chevalier}}]{blair82}
{Blair}, W.~P., {Kirshner}, R.~P., \& {Chevalier}, R.~A. 1982, \apj, 254, 50,
  \dodoi{10.1086/159703}

\bibitem[{{Blair} {et~al.}(1984){Blair}, {Raymond}, {Fesen}, \&
  {Gull}}]{blair84}
{Blair}, W.~P., {Raymond}, J.~C., {Fesen}, R.~A., \& {Gull}, T.~R. 1984, \apj,
  279, 708, \dodoi{10.1086/161936}

\bibitem[{{Blair} {et~al.}(2000){Blair}, {Morse}, {Raymond}, {Kirshner},
  {Hughes}, {Dopita}, {Sutherland}, {Long}, \& {Winkler}}]{blair00}
{Blair}, W.~P., {Morse}, J.~A., {Raymond}, J.~C., {et~al.} 2000, \apj, 537,
  667, \dodoi{10.1086/309077}

\bibitem[{{Braun} \& {Walterbos}(1993{\natexlab{a}})}]{braun93}
{Braun}, R., \& {Walterbos}, R.~A.~M. 1993{\natexlab{a}}, \aaps, 98, 327

\bibitem[{{Braun} \& {Walterbos}(1993{\natexlab{b}})}]{BW}
---. 1993{\natexlab{b}}, \aaps, 98, 327

\bibitem[{{Caldwell} {et~al.}(2009){Caldwell}, {Harding}, {Morrison}, {Rose},
  {Schiavon}, \& {Kriessler}}]{C09}
{Caldwell}, N., {Harding}, P., {Morrison}, H., {et~al.} 2009, \aj, 137, 94,
  \dodoi{10.1088/0004-6256/137/1/94}

\bibitem[{{Chevalier} \& {Kirshner}(1978)}]{chevalier78}
{Chevalier}, R.~A., \& {Kirshner}, R.~P. 1978, \apj, 219, 931,
  \dodoi{10.1086/155855}

\bibitem[{{Chiotellis} {et~al.}(2012){Chiotellis}, {Schure}, \&
  {Vink}}]{chiotellis12}
{Chiotellis}, A., {Schure}, K.~M., \& {Vink}, J. 2012, \aap, 537, A139,
  \dodoi{10.1051/0004-6361/201014754}

\bibitem[{{Conn} {et~al.}(2016){Conn}, {McMonigal}, {Bate}, {Lewis}, {Ibata},
  {Martin}, {McConnachie}, {Ferguson}, {Irwin}, {Elahi}, {Venn}, \&
  {Mackey}}]{Conn}
{Conn}, A.~R., {McMonigal}, B., {Bate}, N.~F., {et~al.} 2016, \mnras, 458,
  3282, \dodoi{10.1093/mnras/stw513}

\bibitem[{{Davidson} {et~al.}(1982){Davidson}, {Walborn}, \&
  {Gull}}]{davidson82}
{Davidson}, K., {Walborn}, N.~R., \& {Gull}, T.~R. 1982, \apjl, 254, L47,
  \dodoi{10.1086/183754}

\bibitem[{{Del Zanna} {et~al.}(2015){Del Zanna}, {Dere}, {Young}, {Landi}, \&
  {Mason}}]{delzanna15}
{Del Zanna}, G., {Dere}, K.~P., {Young}, P.~R., {Landi}, E., \& {Mason}, H.~E.
  2015, \aap, 582, A56, \dodoi{10.1051/0004-6361/201526827}

\bibitem[{{Dickel} \& {Dodorico}(1984)}]{dickel84}
{Dickel}, J.~R., \& {Dodorico}, S. 1984, \mnras, 206, 351,
  \dodoi{10.1093/mnras/206.2.351}

\bibitem[{{Docenko} \& {Sunyaev}(2010)}]{docenko10}
{Docenko}, D., \& {Sunyaev}, R.~A. 2010, \aap, 509, A59,
  \dodoi{10.1051/0004-6361/200810366}

\bibitem[{{Dopita} {et~al.}(2019){Dopita}, {Seitenzahl}, {Sutherland},
  {Nicholls}, {Vogt}, {Ghavamian}, \& {Ruiter}}]{dopita19}
{Dopita}, M.~A., {Seitenzahl}, I.~R., {Sutherland}, R.~S., {et~al.} 2019, \aj,
  157, 50, \dodoi{10.3847/1538-3881/aaf235}

\bibitem[{{Fabricant} {et~al.}(2005){Fabricant}, {Fata}, {Roll}, {Hertz},
  {Caldwell}, {Gauron}, {Geary}, {McLeod}, {Szentgyorgyi}, {Zajac}, {Kurtz},
  {Barberis}, {Bergner}, {Brown}, {Conroy}, {Eng}, {Geller}, {Goddard},
  {Honsa}, {Mueller}, {Mink}, {Ordway}, {Tokarz}, {Woods}, {Wyatt}, {Epps}, \&
  {Dell'Antonio}}]{fab}
{Fabricant}, D., {Fata}, R., {Roll}, J., {et~al.} 2005, \pasp, 117, 1411,
  \dodoi{10.1086/497385}

\bibitem[{{Ghavamian} {et~al.}(2013){Ghavamian}, {Schwartz}, {Mitchell},
  {Masters}, \& {Laming}}]{ghavamian13}
{Ghavamian}, P., {Schwartz}, S.~J., {Mitchell}, J., {Masters}, A., \& {Laming},
  J.~M. 2013, \ssr, 178, 633, \dodoi{10.1007/s11214-013-9999-0}

\bibitem[{{Ghavamian} {et~al.}(2002){Ghavamian}, {Winkler}, {Raymond}, \&
  {Long}}]{ghavamian02}
{Ghavamian}, P., {Winkler}, P.~F., {Raymond}, J.~C., \& {Long}, K.~S. 2002,
  \apj, 572, 888, \dodoi{10.1086/340437}

\bibitem[{{Gordon} {et~al.}(2006){Gordon}, {Bailin}, {Engelbracht}, {Rieke},
  {Misselt}, {Latter}, {Young}, {Ashby}, {Barmby}, {Gibson}, {Hines}, {Hinz},
  {Krause}, {Levine}, {Marleau}, {Noriega-Crespo}, {Stolovy}, {Thilker}, \&
  {Werner}}]{Gordon}
{Gordon}, K.~D., {Bailin}, J., {Engelbracht}, C.~W., {et~al.} 2006, \apjl, 638,
  L87, \dodoi{10.1086/501046}

\bibitem[{{Gull} {et~al.}(2020){Gull}, {Morris}, {Black}, {Nielsen}, {Barlow},
  {Royer}, \& {Swinyard}}]{gull20}
{Gull}, T.~R., {Morris}, P.~W., {Black}, J.~H., {et~al.} 2020, \mnras, 499,
  5269, \dodoi{10.1093/mnras/staa3113}

\bibitem[{{Habing} {et~al.}(1984){Habing}, {Miley}, {Young}, {Baud}, {Boggess},
  {Clegg}, {de Jong}, {Harris}, {Raimond}, {Rowan-Robinson}, \&
  {Soifer}}]{habing}
{Habing}, H.~J., {Miley}, G., {Young}, E., {et~al.} 1984, \apjl, 278, L59,
  \dodoi{10.1086/184223}

\bibitem[{{Hartigan} {et~al.}(1987){Hartigan}, {Raymond}, \&
  {Hartmann}}]{hartigan87}
{Hartigan}, P., {Raymond}, J., \& {Hartmann}, L. 1987, \apj, 316, 323,
  \dodoi{10.1086/165204}

\bibitem[{{Hester} {et~al.}(1994){Hester}, {Raymond}, \& {Blair}}]{hester94}
{Hester}, J.~J., {Raymond}, J.~C., \& {Blair}, W.~P. 1994, \apj, 420, 721,
  \dodoi{10.1086/173598}

\bibitem[{{Hwang} \& {Laming}(2012)}]{hwang12}
{Hwang}, U., \& {Laming}, J.~M. 2012, \apj, 746, 130,
  \dodoi{10.1088/0004-637X/746/2/130}

\bibitem[{{Itoh}(1981)}]{itoh81}
{Itoh}, H. 1981, \pasj, 33, 1

\bibitem[{{Jacobson-Gal{\'a}n} {et~al.}(2020){Jacobson-Gal{\'a}n}, {Margutti},
  {Kilpatrick}, {Hiramatsu}, {Perets}, {Khatami}, {Foley}, {Raymond}, {Yoon},
  {Bobrick}, {Zenati}, {Galbany}, {Andrews}, {Brown}, {Cartier}, {Coppejans},
  {Dimitriadis}, {Dobson}, {Hajela}, {Howell}, {Kuncarayakti}, {Milisavljevic},
  {Rahman}, {Rojas-Bravo}, {Sand}, {Shepherd}, {Smartt}, {Stacey}, {Stroh},
  {Swift}, {Terreran}, {Vinko}, {Wang}, {Anderson}, {Baron}, {Berger},
  {Blanchard}, {Burke}, {Coulter}, {DeMarchi}, {DerKacy}, {Fremling}, {Gomez},
  {Gromadzki}, {Hosseinzadeh}, {Kasen}, {Kriskovics}, {McCully},
  {M{\"u}ller-Bravo}, {Nicholl}, {Ordasi}, {Pellegrino}, {Piro}, {P{\'a}l},
  {Ren}, {Rest}, {Rich}, {Sai}, {S{\'a}rneczky}, {Shen}, {Short}, {Siebert},
  {Stauffer}, {Szak{\'a}ts}, {Zhang}, {Zhang}, \& {Zhang}}]{jacobson-galan20}
{Jacobson-Gal{\'a}n}, W.~V., {Margutti}, R., {Kilpatrick}, C.~D., {et~al.}
  2020, \apj, 898, 166, \dodoi{10.3847/1538-4357/ab9e66}

\bibitem[{{Kaaret}(2002{\natexlab{a}})}]{kaaret}
{Kaaret}, P. 2002{\natexlab{a}}, \apj, 578, 114, \dodoi{10.1086/342475}

\bibitem[{{Kaaret}(2002{\natexlab{b}})}]{kaaret02}
---. 2002{\natexlab{b}}, \apj, 578, 114, \dodoi{10.1086/342475}

\bibitem[{{Kallman} \& {McCray}(1982)}]{kallman82}
{Kallman}, T.~R., \& {McCray}, R. 1982, \apjs, 50, 263, \dodoi{10.1086/190828}

\bibitem[{{Kent}(1989)}]{kent}
{Kent}, S.~M. 1989, \pasp, 101, 489, \dodoi{10.1086/132457}

\bibitem[{{Koo} {et~al.}(2023){Koo}, {Lee}, {Lee}, \& {Yoon}}]{koo23}
{Koo}, B.-C., {Lee}, Y.-H., {Lee}, J.-J., \& {Yoon}, S.-C. 2023, arXiv
  e-prints, arXiv:2305.04484, \dodoi{10.48550/arXiv.2305.04484}

\bibitem[{{Leahy} {et~al.}(2023){Leahy}, {Monaghan}, \& {Ranasinghe}}]{leahy23}
{Leahy}, D., {Monaghan}, C., \& {Ranasinghe}, S. 2023, \aj, 165, 116,
  \dodoi{10.3847/1538-3881/acb68d}

\bibitem[{{Leahy} {et~al.}(2020){Leahy}, {Postma}, {Chen}, \&
  {Buick}}]{leahy20}
{Leahy}, D.~A., {Postma}, J., {Chen}, Y., \& {Buick}, M. 2020, \apjs, 247, 47,
  \dodoi{10.3847/1538-4365/ab77a9}

\bibitem[{{Lee} \& {Lee}(2014)}]{lee14}
{Lee}, J.~H., \& {Lee}, M.~G. 2014, \apj, 786, 130,
  \dodoi{10.1088/0004-637X/786/2/130}

\bibitem[{{Massey} {et~al.}(2016){Massey}, {Neugent}, \& {Smart}}]{Massey2}
{Massey}, P., {Neugent}, K.~F., \& {Smart}, B.~M. 2016, \aj, 152, 62,
  \dodoi{10.3847/0004-6256/152/3/62}

\bibitem[{{Massey} {et~al.}(2006){Massey}, {Olsen}, {Hodge}, {Strong},
  {Jacoby}, {Schlingman}, \& {Smith}}]{M06}
{Massey}, P., {Olsen}, K.~A.~G., {Hodge}, P.~W., {et~al.} 2006, \aj, 131, 2478,
  \dodoi{10.1086/503256}

\bibitem[{{Medina} {et~al.}(2014){Medina}, {Raymond}, {Edgar}, {Caldwell},
  {Fesen}, \& {Milisavljevic}}]{medina14}
{Medina}, A.~A., {Raymond}, J.~C., {Edgar}, R.~J., {et~al.} 2014, \apj, 791,
  30, \dodoi{10.1088/0004-637X/791/1/30}

\bibitem[{{Miko{\l}ajewska} {et~al.}(2014){Miko{\l}ajewska}, {Caldwell}, \&
  {Shara}}]{miko}
{Miko{\l}ajewska}, J., {Caldwell}, N., \& {Shara}, M.~M. 2014, \mnras, 444,
  586, \dodoi{10.1093/mnras/stu1480}

\bibitem[{{Milisavljevic} {et~al.}(2015){Milisavljevic}, {Margutti}, {Kamble},
  {Patnaude}, {Raymond}, {Eldridge}, {Fong}, {Bietenholz}, {Challis},
  {Chornock}, {Drout}, {Fransson}, {Fesen}, {Grindlay}, {Kirshner}, {Lunnan},
  {Mackey}, {Miller}, {Parrent}, {Sanders}, {Soderberg}, \&
  {Zauderer}}]{milisavljevic15}
{Milisavljevic}, D., {Margutti}, R., {Kamble}, A., {et~al.} 2015, \apj, 815,
  120, \dodoi{10.1088/0004-637X/815/2/120}

\bibitem[{{Millard} {et~al.}(2021){Millard}, {Ravi}, {Rho}, \&
  {Park}}]{millard21}
{Millard}, M.~J., {Ravi}, A.~P., {Rho}, J., \& {Park}, S. 2021, \apjs, 257, 36,
  \dodoi{10.3847/1538-4365/ac1d4a}

\bibitem[{{Morris} {et~al.}(2017){Morris}, {Gull}, {Hillier}, {Barlow},
  {Royer}, {Nielsen}, {Black}, \& {Swinyard}}]{morris17}
{Morris}, P.~W., {Gull}, T.~R., {Hillier}, D.~J., {et~al.} 2017, \apj, 842, 79,
  \dodoi{10.3847/1538-4357/aa71b3}

\bibitem[{{Morse} {et~al.}(2001){Morse}, {Kellogg}, {Bally}, {Davidson},
  {Balick}, \& {Ebbets}}]{morse01}
{Morse}, J.~A., {Kellogg}, J.~R., {Bally}, J., {et~al.} 2001, \apjl, 548, L207,
  \dodoi{10.1086/319092}

\bibitem[{{Narita} {et~al.}(2023){Narita}, {Uchida}, {Yoshida}, {Tanaka}, \&
  {Tsuru}}]{narita23}
{Narita}, T., {Uchida}, H., {Yoshida}, T., {Tanaka}, T., \& {Tsuru}, T.~G.
  2023, arXiv e-prints, arXiv:2304.11819, \dodoi{10.48550/arXiv.2304.11819}

\bibitem[{{National Optical Astronomy Observatories}(1999)}]{iraf}
{National Optical Astronomy Observatories}. 1999, {IRAF: Image Reduction and
  Analysis Facility}, Astrophysics Source Code Library, record ascl:9911.002.
\newblock \doeprint{9911.002}

\bibitem[{{Neugent} {et~al.}(2012){Neugent}, {Massey}, \& {Georgy}}]{neugent}
{Neugent}, K.~F., {Massey}, P., \& {Georgy}, C. 2012, \apj, 759, 11,
  \dodoi{10.1088/0004-637X/759/1/11}

\bibitem[{{Raymond}(1979)}]{raymond79}
{Raymond}, J.~C. 1979, \apjs, 39, 1, \dodoi{10.1086/190562}

\bibitem[{{Raymond}(2018)}]{raymond18}
---. 2018, \ssr, 214, 28, \dodoi{10.1007/s11214-017-0453-6}

\bibitem[{{Raymond} {et~al.}(2020){Raymond}, {Chilingarian}, {Blair},
  {Sankrit}, {Slavin}, \& {Burkhart}}]{raymond20}
{Raymond}, J.~C., {Chilingarian}, I.~V., {Blair}, W.~P., {et~al.} 2020, \apj,
  894, 108, \dodoi{10.3847/1538-4357/ab886d}

\bibitem[{{Raymond} {et~al.}(2023){Raymond}, {Ghavamian}, {Bohdan}, {Ryu},
  {Niemiec}, {Sironi}, {Tran}, {Amato}, {Hoshino}, {Pohl}, {Amano}, \&
  {Fiuza}}]{raymond23}
{Raymond}, J.~C., {Ghavamian}, P., {Bohdan}, A., {et~al.} 2023, arXiv e-prints,
  arXiv:2303.08849, \dodoi{10.48550/arXiv.2303.08849}

\bibitem[{{Sanders} {et~al.}(2012){Sanders}, {Caldwell}, {McDowell}, \&
  {Harding}}]{sanders}
{Sanders}, N.~E., {Caldwell}, N., {McDowell}, J., \& {Harding}, P. 2012, \apj,
  758, 133, \dodoi{10.1088/0004-637X/758/2/133}

\bibitem[{{Sasaki} {et~al.}(2012){Sasaki}, {Pietsch}, {Haberl},
  {Hatzidimitriou}, {Stiele}, {Williams}, {Kong}, \& {Kolb}}]{sasaki12}
{Sasaki}, M., {Pietsch}, W., {Haberl}, F., {et~al.} 2012, \aap, 544, A144,
  \dodoi{10.1051/0004-6361/201219025}

\bibitem[{{Sasaki} {et~al.}(2018){Sasaki}, {Haberl}, {Henze}, {Saeedi},
  {Williams}, {Plucinsky}, {Hatzidimitriou}, {Karampelas}, {Sokolovsky},
  {Breitschwerdt}, {de Avillez}, {Filipovi{\'c}}, {Galvin}, {Kavanagh}, \&
  {Long}}]{sasaki18}
{Sasaki}, M., {Haberl}, F., {Henze}, M., {et~al.} 2018, \aap, 620, A28,
  \dodoi{10.1051/0004-6361/201833588}

\bibitem[{{Schlafly} \& {Finkbeiner}(2011)}]{schlafly}
{Schlafly}, E.~F., \& {Finkbeiner}, D.~P. 2011, \apj, 737, 103,
  \dodoi{10.1088/0004-637X/737/2/103}

\bibitem[{{Sibley} {et~al.}(2016){Sibley}, {Katz}, {Satterfield}, {Vanderveer},
  \& {MacAlpine}}]{sibley16}
{Sibley}, A.~R., {Katz}, A.~M., {Satterfield}, T.~J., {Vanderveer}, S.~J., \&
  {MacAlpine}, G.~M. 2016, \aj, 152, 93, \dodoi{10.3847/0004-6256/152/4/93}

\bibitem[{{Smith} {et~al.}(2003){Smith}, {Gehrz}, {Hinz}, {Hoffmann}, {Hora},
  {Mamajek}, \& {Meyer}}]{smith03}
{Smith}, N., {Gehrz}, R.~D., {Hinz}, P.~M., {et~al.} 2003, \aj, 125, 1458,
  \dodoi{10.1086/346278}

\bibitem[{{Smith} \& {Morse}(2004)}]{smith04}
{Smith}, N., \& {Morse}, J.~A. 2004, \apj, 605, 854, \dodoi{10.1086/382671}

\bibitem[{{Smith} {et~al.}(2018){Smith}, {Andrews}, {Rest}, {Bianco}, {Prieto},
  {Matheson}, {James}, {Smith}, {Strampelli}, \& {Zenteno}}]{smith18}
{Smith}, N., {Andrews}, J.~E., {Rest}, A., {et~al.} 2018, \mnras, 480, 1466,
  \dodoi{10.1093/mnras/sty1500}

\bibitem[{{Stiele} {et~al.}(2011){Stiele}, {Pietsch}, {Haberl},
  {Hatzidimitriou}, {Barnard}, {Williams}, {Kong}, \& {Kolb}}]{stiele11}
{Stiele}, H., {Pietsch}, W., {Haberl}, F., {et~al.} 2011, \aap, 534, A55,
  \dodoi{10.1051/0004-6361/201015270}

\bibitem[{{Sutherland} \& {Dopita}(1995)}]{sutherland95}
{Sutherland}, R.~S., \& {Dopita}, M.~A. 1995, \apj, 439, 381,
  \dodoi{10.1086/175181}

\bibitem[{{Temim} {et~al.}(2022){Temim}, {Slane}, {Raymond}, {Patnaude},
  {Murray}, {Ghavamian}, {Renzo}, \& {Jacovich}}]{temim22}
{Temim}, T., {Slane}, P., {Raymond}, J.~C., {et~al.} 2022, \apj, 932, 26,
  \dodoi{10.3847/1538-4357/ac6bf4}

\bibitem[{{Walterbos} \& {Braun}(1992)}]{WB}
{Walterbos}, R.~A.~M., \& {Braun}, R. 1992, \aaps, 92, 625

\bibitem[{{Weil} {et~al.}(2020){Weil}, {Fesen}, {Patnaude}, {Raymond},
  {Chevalier}, {Milisavljevic}, \& {Gerardy}}]{weil20}
{Weil}, K.~E., {Fesen}, R.~A., {Patnaude}, D.~J., {et~al.} 2020, \apj, 891,
  116, \dodoi{10.3847/1538-4357/ab76bf}

\bibitem[{{Wenger} {et~al.}(2000){Wenger}, {Ochsenbein}, {Egret}, {Dubois},
  {Bonnarel}, {Borde}, {Genova}, {Jasniewicz}, {Lalo{\"e}}, {Lesteven}, \&
  {Monier}}]{simbad}
{Wenger}, M., {Ochsenbein}, F., {Egret}, D., {et~al.} 2000, \aaps, 143, 9,
  \dodoi{10.1051/aas:2000332}

\bibitem[{{Winkler} {et~al.}(2017){Winkler}, {Blair}, \& {Long}}]{winkler17}
{Winkler}, P.~F., {Blair}, W.~P., \& {Long}, K.~S. 2017, \apj, 839, 83,
  \dodoi{10.3847/1538-4357/aa683d}

\end{thebibliography}
\bibliographystyle{aasjournal}



\end{document}